\pdfoutput=1

\documentclass[11pt]{article}

\usepackage[]{acl}

\usepackage{times}
\usepackage{latexsym}
\usepackage{multirow}
\usepackage[T1]{fontenc}

\usepackage[utf8]{inputenc}
\usepackage{microtype}
\usepackage{inconsolata}
\usepackage{url}
\usepackage{amssymb}
\usepackage{graphicx}
\usepackage{booktabs}
\usepackage{CJKutf8}
\usepackage{graphicx}
\usepackage{subfigure}
\usepackage{float}
\usepackage{amsmath}
\usepackage{amssymb}
\usepackage{multirow}
\usepackage{booktabs}
\usepackage{url}
\usepackage{array}
\usepackage{enumitem}
\usepackage{algorithm}
\usepackage{algorithmic}
\usepackage{pifont}
\usepackage{bm}
\usepackage{lipsum}
\usepackage{xcolor}
\usepackage{makecell}
\usepackage{CJKutf8}
\usepackage{tikz}
\usepackage[edges]{forest}
\usepackage{array}
\usepackage{hyperref} 
\usepackage{soul}
\sethlcolor{yellow}
\soulregister\cite7
\soulregister\ref7

\definecolor{hidden-pink}{RGB}{255,245,247}
\definecolor{hidden-draw}{RGB}{20,68,106}

\title{Agents in Software Engineering: Survey, Landscape, and Vision}
\author{Yanlin Wang$^1$, Wanjun Zhong$^1$, Yanxian Huang$^1$\thanks{*Corresponding author: Yanxian Huang, \url{huangyx353@mail2.sysu.edu.cn}} , Ensheng Shi$^2$, Min Yang$^3$, \\ \textbf{Jiachi Chen$^1$, Hui Li$^4$, Yuchi Ma$^5$, Qianxiang Wang$^5$, Zibin Zheng$^1$}\\ 
   $^1$Sun Yat-sen University, $^2$Xi’an Jiaotong University \\
   $^3$Shenzhen Institute of Advanced Technology, Chinese Academy of Sciences \\
   $^4$Xiamen University, $^5$Huawei Cloud Computing Technologies Co., Ltd.}

\usepackage{xspace}

\newcommand{\sect}{Section\xspace}

\begin{document}
\maketitle
\begin{abstract}
In recent years, Large Language Models (LLMs) have achieved remarkable success and have been widely used in various downstream tasks, especially in the tasks of the software engineering (SE) field. We find that many studies combining LLMs with SE have employed the concept of agents either explicitly or implicitly. However, there is a lack of an in-depth survey to sort out the development context of existing works, analyze how existing works combine the LLM-based agent technologies to optimize various tasks, and clarify the framework of LLM-based agents in SE. 
In this paper, we conduct the first survey of the studies on combining LLM-based agents with SE and present a framework of LLM-based agents in SE which includes three key modules: perception, memory, and action. 
We also summarize the current challenges in combining the two fields and propose future opportunities in response to existing challenges. We maintain a GitHub repository of the related papers at: \url{https://github.com/DeepSoftwareAnalytics/Awesome-Agent4SE}.
\end{abstract}

\section{Introduction}

In recent years, Large Language Models (LLMs) have achieved remarkable success and have been widely used in many downstream tasks, especially in various tasks in the field of software engineering (SE)~\cite{zheng2308towards}, such as code summarization~\cite{ahmed2024automatic, sun2023automatic, haldar2024analyzing, mao2024automated, guo2023snippet, wang2021cocosum}, code generation~\cite{Jiang2023SelfplanningCG, Hu2024LeveragingPD, Yang2023ChainofThoughtIN, Tian2023TestCaseDrivenPU, Li2023ThinkOT, wang2024teaching}, code translation~\cite{pan2024lost, pan2308understanding}, vulnerability detection and repair~\cite{zhou2024large, islam2024code, de2024enhanced, le2024study, liu2024prompt, chen2023chatgpt}, etc. Many studies combining LLMs with SE have employed the concept of agents from the artificial intelligence field, either explicitly or implicitly. Explicit use indicates that the paper directly mentions the use of agent-related technologies, whereas implicit use suggests that while the concept of intelligent agents is utilized, it may be described using different terminology or presented in alternative forms.

An agent~\cite{wang2024survey} represents an intelligent entity capable of perceiving, reasoning, and taking action. It perceives the environment's state and selects actions based on its goals and design to maximize specific performance metrics, serving as a crucial technical foundation for accomplishing diverse tasks and objectives. LLM-based agents generally use LLMs as the cognitive core of the agent and perform well in scenarios such as automation, intelligent control, and human-computer interaction, leveraging the powerful capabilities of LLMs in language understanding and generation, learning and reasoning, context perception and memory, and multimodality, etc.
With the development of various fields, the concept of traditional and LLMs-based agents has gradually become clear and widely used in the field of Natural Language Processing (NLP)~\cite{ xi2023rise}. However, although existing works either explicitly or implicitly use this concept in SE, there is still no clear definition of agents. There is a lack of an in-depth survey to analyze how existing works combine the agent technologies to optimize various tasks, sort out the development context of existing works, and clarify the framework of agents in SE.

In this paper, we conduct an in-depth analysis of the work on combining LLM-based agents with SE, summarize the current challenges in combining the two fields, and propose possible opportunities for future research in response to existing challenges. Specifically, we first collect papers on the application of LLM-based agent technology to SE and obtain 115 papers after filtering and quality assessment.
Then, inspired by the definition of the traditional agent~\cite{wang2024survey, xi2023rise}, we present a general conceptual framework for the LLM-based agents in SE (\sect~\ref{sec:agent4SE}), comprising three key components: perception, memory, action. We first introduce the perception module (\sect~\ref{sec:perception}), which can handle inputs of different modalities, such as textual input, visual input, auditory input, etc. Next, we present the memory module(\sect~\ref{sec:memory}), which includes semantic memory, episodic memory, and procedural memory, helping the agent to make reasoning decisions. Finally, we introduce the action module(\sect~\ref{sec:action}) which contains internal actions such as reasoning, retrieval, and learning, as well as external actions like interacting with the environment.
After that, we provide a detailed and thorough introduction to the challenges and opportunities of LLM-based agents in SE (\sect~\ref{sec:challenges&opportunities}). Specifically, we propose the following opportunities for future research in response to the current challenges of LLM-based agents in SE:

\begin{itemize}
\item  Most existing work on exploring perception modules focuses on token-based input in textual input but lacks work on exploring other modalities. 
\item Numerous new tasks remain outside the current learning scope of LLMs, and complex tasks in the SE field necessitate agents with a diverse range of capabilities. Therefore, it is crucial to explore how LLM-based agents play new roles and effectively balance the ability to perform multiple roles.
\item It lacks an authoritative and recognized knowledge base containing rich code-related knowledge as an external retrieval base in the SE field.
\item Alleviating the hallucinations of LLM-based agents can improve the overall performance of the agent, while agent optimization can reversely alleviate the hallucinations of LLM-based agents.
\item The multi-agent collaboration process requires a large amount of computing resources and additional communication overhead generated by synchronizing and sharing various types of information. Exploring technologies to improve the efficiency of multi-agent collaboration is also an opportunity for future work.
\item Technologies in the SE field can also promote the development of the Agent field, needing more work to explore integrating more advanced technologies in the SE field into Agent, promoting the development of Agent and progress in the SE field.
\end{itemize}

In addition, The technologies in SE, especially those related to code, can also promote the development of the agent field, indicating the mutually reinforcing relationship between these two very different fields. However, there is little work exploring SE technology in agents, and the focus is still on the simple and basic technology of SE, such as function calls, HTTP requests, and other tools. Therefore, in this paper, we mainly focus on the work related to agents in SE, and only briefly discuss the studies about SE technology in agents in Section~\ref{sec:se4agent}, as an opportunity for future work.

\tikzstyle{my-box}=[
    rectangle,
    draw=hidden-draw,
    rounded corners,
    text opacity=1,
    minimum height=1.5em,
    minimum width=5em,
    inner sep=2pt,
    align=center,
    fill opacity=.5,
    line width=0.8pt,
]
\tikzstyle{leaf}=[my-box, minimum height=1.5em,
    fill=hidden-pink!80, text=black, align=center,font=\LARGE,
    inner xsep=2pt,
    inner ysep=4pt,
    line width=0.8pt,
]
\begin{figure*}[t!]
    \centering
    \resizebox{1.05\textwidth}{!}{
        \begin{forest}
            forked edges,
            for tree={
                grow=east,
                reversed=true,
                anchor=base west,
                parent anchor=east,
                child anchor=west,
                base=center,
                font=\huge,
                rectangle,
                draw=hidden-draw,
                rounded corners,
                align=center,
                text centered,
                minimum width=5em,
                edge+={darkgray, line width=1pt},
                s sep=8pt,
                inner xsep=2pt,
                inner ysep=3pt,
                line width=0.8pt,
                ver/.style={rotate=90, child anchor=north, parent anchor=south, anchor=center},
            },
            where level=1{text width=10em, font=\huge,}{},
            where level=2{text width=28em, font=\huge,}{},
            where level=3{text width=23em,font=\huge,}{},
            where level=4{text width=23em,font=\huge,}{},
            [
                Agents in Software Engineering, ver
                [
                    Perception\\ (\S \ref{sec:perception}),
                    [
                        Textual Input
                        [
                            Token-based Input
                            [
                                \citet{ahmed2024automatic}{,}
                                \citet{alkaswan2023extending}{,}\\
                                \citet{arakelyan2023exploring}{, }
                                \citet{Alqarni2022LowLS}{, } \\
                                \citet{Beurer_Kellner_2023}, leaf, text width=35em
                            ]
                        ]
                        [
                            Tree/graph-based Input
                            [
                                \citet{Ma2023BridgingCS}{, }
                                \citet{Ma2023LMsUC}{,}\\
                                \citet{Zhang2023NeuralPR}{,}
                                \citet{10123540}, leaf, text width=35em
                            ]   
                        ]
                        [
                            Hybrid-based Input
                            [
                                \citet{Niu2022SPTCodeSP}, leaf, text width=35em
                            ]
                        ]
                    ]
                    [
                        Visual Input 
                        [
                            \citet{Behrang2018GUIFetchSA}{,}
                            \citet{Reiss2014SeekingTU}{,} \\
                            \citet{Xie2019UserIC}{,}
                            \citet{Driess2023PaLMEAE},leaf, text width=33em
                        ]
                    ]
                    [
                        Auditory Input 
                        [
                            \citet{Bao2020psc2code}, leaf, text width=33em
                        ]
                    ]
                ]
                [
                    Memory\\ (\S \ref{sec:memory})
                    [
                        {Semantic Memory}\\{(Mainly existing in the external}\\{knowledge bases, including doc,}\\ {libraries, API information)}
                        [
                            \citet{wang2024teaching}{,}
                            \citet{Zhang2024CodeAgentEC}{,}\\
                            \citet{Eghbali2024DeHallucinatorIG}{,}
                            \citet{Patel2023EvaluatingIL}{,}\\
                            \citet{Zhou2022DocPromptingGC}{,}
                            \citet{Ren2023FromMT}{,}\\
                            \citet{Sen2023CriticalDD}{,}
                            \citet{Zhang2023ToolCoderTC}, leaf, text width=33em
                        ]
                    ]
                    [
                        {Episodic Memory}\\{Mainly including history messages,}\\ {retrieving relevant codes from}\\ {codebase, examples involved in }\\{ICL technology, etc} 
                        [
                            \citet{Li2023AceCoderUE}{,}
                            \citet{Ren2023FromMT}{,}\\
                            \citet{Wei2023CoeditorLC}{,}
                            \citet{Zhang2023RepoCoderRC}{,}\\
                            \citet{Eghbali2024DeHallucinatorIG}{,}
                            \citet{Ahmed2023AutomaticSA}{,}\\
                            \citet{Feng2023PromptingIA}, leaf, text width=33em
                        ]
                    ]
                    [
                        Procedural Memory
                        [
                            {Implicit Knowledge}\\{(Stored in the LLM weights)}
                            [
                                \citet{christopoulou2022pangu}
                                \citet{shen2023pangu}\\
                                \citet{gunasekar2023textbooks}
                                \citet{fried2022incoder}\\
                                \citet{roziere2023code}
                                \citet{zan2022cert}\\
                                \citet{xu2023wizardlm}
                                \citet{thoppilan2022lamda}\\
                                \citet{chandel2022training}
                                \citet{li2022competition}\\
                                \citet{luo2023wizardcoder}
                                \citet{Svyatkovskiy2020IntelliCodeCC}{,}\\
                                \citet{Ahmad2021UnifiedPF}{,}
                                \citet{Christopoulou2022PanGuCoderPS}, leaf, text width=34em
                            ]    
                        ]
                        [
                            {Explicit Knowledge}\\{(Written in the agent’s code)}
                            [
                                \citet{Patel2023EvaluatingIL} {,}
                                \citet{Shin2023PromptEO}{,}\\
                                \citet{Zhang2023PromptEnhancedSV}{,}
                                \citet{Zan2022CERTCP}{,}\\
                                \citet{Xu2023WizardLMEL}{,}
                                \citet{Thoppilan2022LaMDALM}, leaf, text width=34em
                            ]
                        ]
                    ]
                ]
                [
                    Action \\(\S \ref{sec:action})
                    [
                        Internal Actions
                        [
                            Reasoning Actions
                            [
                                \citet{inproceedings}{,}
                                \citet{Feng2023PromptingIA}{,}\\
                                \citet{Jiang2023SelfplanningCG}{,}
                                \citet{Bairi2023CodePlanRC}{,}\\
                                \citet{Li2023StructuredCP}{,}
                                \citet{huang2024codecot}{,}\\
                                \citet{Li2023ThinkOT}{,}
                                \citet{Tian2023TestCaseDrivenPU}{,}\\
                                \citet{Christianos2023PanguAgentAF}{,}
                                \citet{Zhang2023DraftV}{,}\\
                                \citet{Yang2023ChainofThoughtIN}{,}
                                \citet{Wang2024TeachingCL}{,}\\
                                \citet{Hu2024LeveragingPD}{,}
                                \citet{Le2023CodeChainTM}{,}\\
                                \citet{Ma2023BridgingCS}, leaf, text width=34em
                            ]
                        ]
                        [
                            Retrieval Actions
                            [
                                \citet{Zan2022WhenLM}{,}
                                \citet{Zhang2023RepoCoderRC}{,}\\
                                \citet{Li2022GenerationAugmentedQE}{,}
                                \citet{Li2023LargeLM}{,}\\
                                \citet{Nashid2023RetrievalBasedPS}{,}
                                \citet{Zhang2024CodeAgentEC}{,}\\
                                \citet{Geng2023LargeLM}{,}
                                \citet{Zhou2022DocPromptingGC}{,}\\
                                \citet{Li2023AceCoderUE}{,}
                                \citet{Eghbali2024DeHallucinatorIG}{,}\\
                                \citet{Zhang2023ToolCoderTC}{,}
                                \citet{Xia2023ThePS}, leaf, text width=34em
                            ]
                        ]
                        [
                            Learning Actions
                            [
                                Updating Implicit Knowledge
                                [
                                    \citet{Weyssow2023ExploringPF}{,}
                                    \citet{Xia2023RevisitingTP}{,}\\
                                    \citet{Wang2023OneAF}{,}
                                    \citet{Wei2023CoeditorLC}{,}\\
                                    \citet{Wei2023MagicoderSC}, leaf, text width=30em
                                ]
                            ]
                            [
                                Updating Semantic Memory \\with Knowledge
                                [
                                    \citet{Hu2023InstructCoderEL}{,}
                                    \citet{Zan2023CanPL}{,}\\
                                    \citet{Wu2022SelfAdaptiveIL}{,}
                                    \citet{Muennighoff2023OctoPackIT}, leaf, text width=30em
                                ]
                            ]
                            [
                                Updating Agent Code
                                [
                                    \citet{Li2023CodeIELC}, leaf, text width=30em
                                ]
                            ]
                        ]
                    ]
                    [
                        External Actions
                        [
                            Dialogue with Human/Agent
                            [
                                \citet{Jain2023CoarseTuningMO}{,}
                                \citet{Paul2023REFINERRF}{,}\\
                                \citet{Yang2023SupervisedKM}{,}
                                \citet{Mu2023ClarifyGPTEL}{,}\\
                                \citet{Moon2023CoffeeBY} {,}
                                \citet{Shojaee2023ExecutionbasedCG}{,}\\
                                \citet{Liu2023RLTFRL}{,}
                                \citet{Wang2023ChatCoderCR}{,}\\
                                \citet{Sun2023CloverCV}{,}
                                \citet{Shinn2023ReflexionLA}{,} \\
                                \citet{Madaan2023SelfRefineIR}{,}
                                \citet{Wei2023CopilotingTC}{,}\\
                                \citet{Wang2023MINTEL}{,}
                                \citet{Hong2023MetaGPTMP}{,}\\
                                \citet{Huang2023AgentCoderMC}, leaf, text width=34em
                            ]
                        ]
                        [
                            Digital Environments
                            [
                                \citet{Wei2023CopilotingTC}{,}
                                \citet{Agrawal2023GuidingLM}{,}\\
                                \citet{Zhang2024CodeAgentEC}{,}
                                \citet{Wang2024TeachingCL}{,}\\
                                \citet{Zhang2023SelfEditFC}{,}
                                \citet{Wang2023MINTEL}{,}\\
                                \citet{Jain2023CoarseTuningMO}{,}
                                \citet{Shojaee2023ExecutionbasedCG} {,}\\
                                \citet{Liu2023RLTFRL} {,}
                                \citet{Shinn2023ReflexionLA}{,}\\
                                \citet{Wang2022CompilableNC}, leaf, text width=34em
                            ]
                        ]
                    ]
                ]
            ]
        \end{forest}
    }
    \caption{Taxonomy of LLM-based agents in software engineering.}
    \label{fig:taxonomy}
\end{figure*}
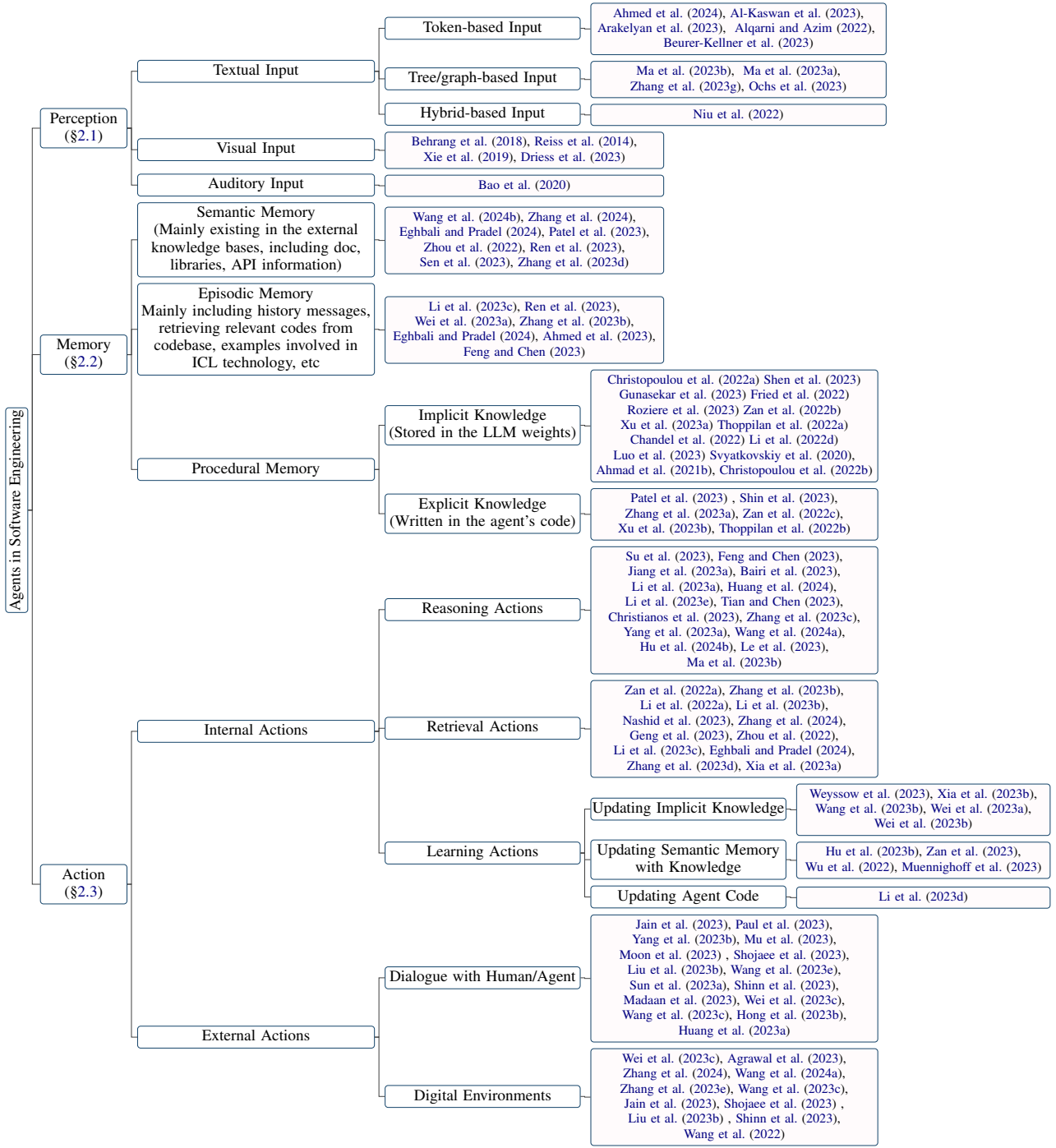

\section{LLM-based Agents in SE}
\label{sec:agent4SE}
\begin{figure*}[t]
\centering
\includegraphics[width=\linewidth]{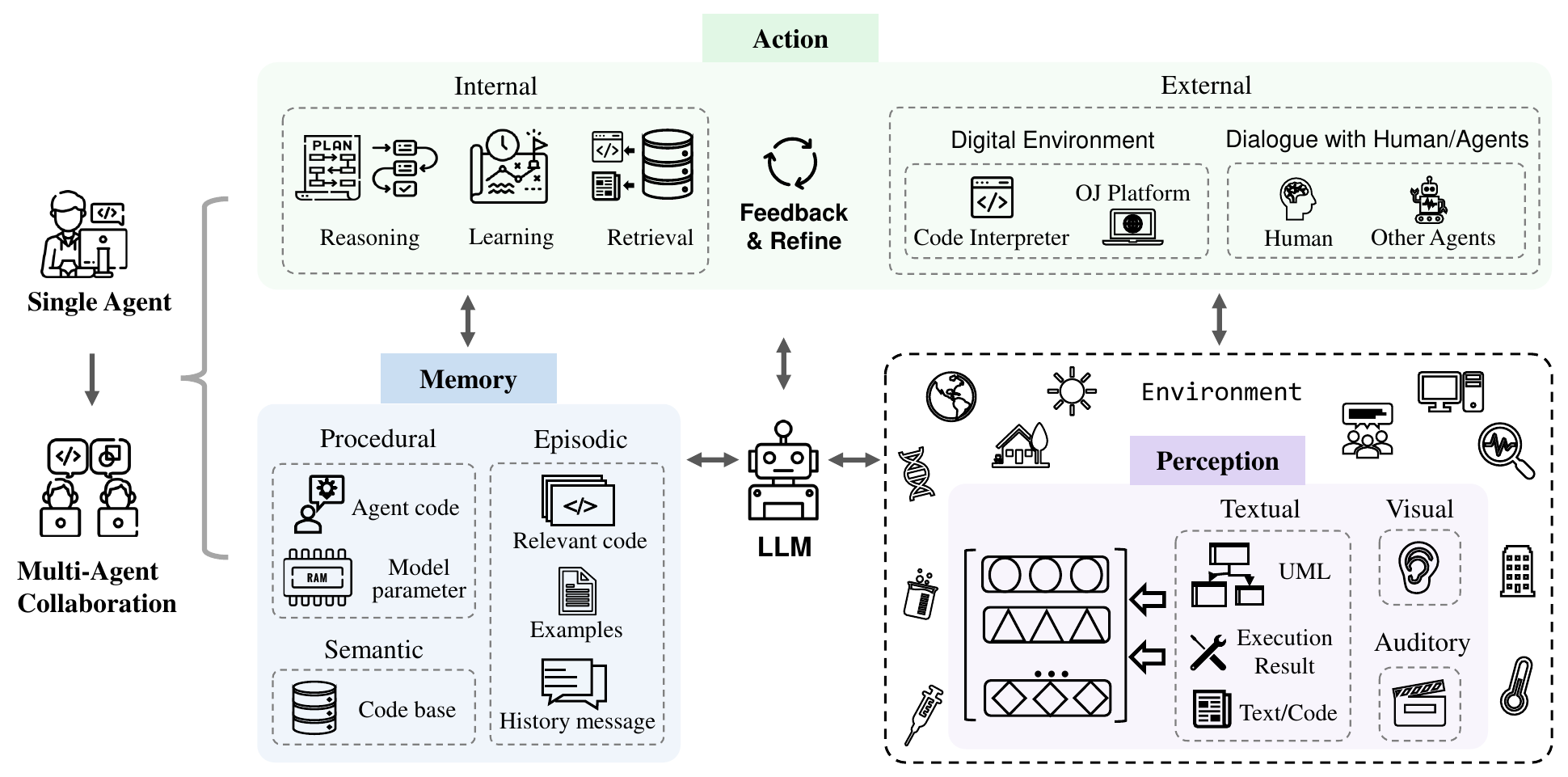}
\caption{An overview of agent framework in SE.}
\label{fig:framework}
\end{figure*}

We present a framework of the LLM-based agents in SE after sorting out and analyzing the studies obtained during data collection. As shown in Figure~\ref{fig:framework}, a single agent contains three key modules: perception, memory, and action. Specifically, the perception module receives external environment information of various modalities and converts it into an input form that the LLM can understand and process. The action module includes internal and external actions, which are responsible for making reasoning decisions based on the input of LLM and refining the decisions based on the feedback obtained from interacting with the external environment, respectively. The memory module includes semantic, episodic, and procedural memory, which can provide additional useful information to help LLM make reasoning decisions. At the same time, the action module can also update different memories in the memory module by learning actions, providing more effective memory information for reasoning and retrieval actions. Furthermore, multi-agent collaboration consists of multiple single agents, who are responsible for part of the task and complete the task together through collaborative cooperation.
In this section, we will introduce the details of each module in the framework of the LLM-based agents in SE.

\subsection{Perception}
\label{sec:perception}
The perception module connects the LLM-based agent to the external environment and is the core of processing external input.
It can process inputs of different modalities such as textual, visual, and auditory input, and convert them into an embedding format that LLM-based agents can understand and process, laying the foundation for reasoning and decision-making actions of LLM-based agents. Next, we will introduce the details of different modal inputs in the perception module.

\subsubsection{Textual Input}
Different from the textual input format in NLP, considering the characteristics of code, the textual input format in the SE includes token-based, tree/graph-based, and hybrid-based input.

\textbf{Token-based Input.} Token-based input~\citep{ahmed2024automatic, alkaswan2023extending, arakelyan2023exploring, Beurer_Kellner_2023, Alqarni2022LowLS, li2022exploring, gu2022accelerating, du2021single} is the most mainstream input mode, which directly regards the code as natural language text and directly uses the token sequence as the input of LLM, ignoring the characteristics of code.

\textbf{Tree/Graph-based Input.}
Compared to natural language, code has strict structure and grammatical rules, and is usually written following the grammar of a specific programming language. Based on the characteristics of code, tree/graph-based input~\citep{Ma2023BridgingCS, Ma2023LMsUC, Zhang2023NeuralPR, 10123540, bi2024make, shi2023cocoast, shi2021cast, wang2021code} can convert code into tree structures such as abstract syntax trees or graph structures like control flow graphs to model the structural information of code. However, there is a challenge and opportunity that current work related to LLM-based SE agents has not explored such modalities.

\textbf{Hybrid-based Input.}
Hybrid input~\citep{Niu2022SPTCodeSP, hu2024tackling, guo2022unixcoder} combines multiple modalities to provide LLM with different types of information. For example, hybrid input containing token-based and tree-based input can combine the semantic and structural information of the code to better model and understand the code. However, there is also no work related to LLM-based agents in SE exploring this modality.

\subsubsection{Visual Input}
Visual input uses visual image data such as UI sketches or UML design drawings as model input and makes inference decisions through modeling and analysis of images. Many related works in NLP have explored this modality. For example,~\citet{Driess2023PaLMEAE} propose PaLM-E, an embodied multi-modal language model whose inputs are multi-modal sentences that interleave visual, continuous state estimation, and textual input encodings. Traditional soft engineering fields also have tasks for visualizing input, such as UI code search~\citep{Behrang2018GUIFetchSA, Reiss2014SeekingTU, Xie2019UserIC} which uses UI sketches as queries for useful code snippets. However, there is a lack of work on visual modality as inputs to LLMs.

\subsubsection{Auditory Input}
Auditory input takes auditory data such as audio as input and interacts with LLM in the form of speech. Traditional software engineering fields have tasks for auditory input, such as programming video search~\citep{Bao2020psc2code} which uses videos as sources as sources for useful code snippets. However, there is also a lack of work on auditory input for LLMs.

\subsection{Memory}
\label{sec:memory}
The memory modules include semantic, episodic, and procedural memory, which can provide additional useful information to help LLM make reasoning decisions. Next, we will introduce the details of these three types of memory respectively.

\subsubsection{Semantic Memory}
Semantic memory stores acknowledged world knowledge of LLM-based agents, usually in the form of external knowledge retrieval bases which include documents, libraries, APIs, or other knowledge.  There have been many works~\cite{wang2024teaching, Zhang2024CodeAgentEC, Eghbali2024DeHallucinatorIG, Patel2023EvaluatingIL,Zhou2022DocPromptingGC, Ren2023FromMT,Zhang2023ToolCoderTC} exploring semantic memory.
Specifically, documents and APIs are the most common information in external knowledge bases. 
For example,~\citet{Zhou2022DocPromptingGC} introduce a novel natural-language-to-code generation approach named DocPrompting, which explicitly leverages documentation by retrieving the relevant documentation pieces based on an NL intent.~\citet{Zhang2024CodeAgentEC} constructs a manually curated benchmark for repo-level code generation named CODEAGENTBENCH, which contains documentation, code dependency, and runtime environment information.~\citet{Ren2023FromMT} propose KPC, a novel Knowledge-driven Prompt Chaining-based code generation approach, which utilizes fine-grained exception-handling knowledge extracted from API documentation to assist LLMs in code generation.
In addition to documents, APIs are also common information in external knowledge bases. For example,~\citet{Eghbali2024DeHallucinatorIG} propose De-Hallucinator, an LLM-based code completion technique, which automatically identifies project-specific API references related to code prefixes and the model's initial predictions, and adds these
referenced information to the prompt.~\citet{Zhang2023ToolCoderTC} integrate API search tools into the generation process, allowing the model to select an API  automatically using the search tool to get suggestions.
In addition, some works also involve other information. For example,~\citet{Patel2023EvaluatingIL} examines the capabilities and limitations of different LLMs in generating code based on libraries defined in context.
~\citet{wang2024teaching} uses augmented functions, along with their corresponding docstrings, to fine-tune a selected code LLM.

\subsubsection{Episodic Memory}
Episodic memory records content related to the current case and experience information from previous decision-making processes. Content related to the current case (such as relevant information found in the search database, samples provided by In-context learning (ICL) technology, etc.) can provide additional knowledge for LLM reasoning, so many works introduce such information into the reasoning process of LLM~\cite{zhong2024memorybank, Li2023AceCoderUE, Feng2023PromptingIA, Ahmed2023AutomaticSA, Wei2023CoeditorLC, Ren2023FromMT, Zhang2023RepoCoderRC, Eghbali2024DeHallucinatorIG, shi2022race}. For example,~\citet{Li2023AceCoderUE} propose a new prompting technique named AceCoder, which selects similar programs as examples in prompts. It provides lots of relevant content (e.g., algorithms, APIs) about the target code.~\citet{Feng2023PromptingIA} propose AdbGPT, a new lightweight approach without any training and hard-coding effort, which can automatically reproduce the bugs from bug reports using In-context learning techniques.~\citet{Ahmed2023AutomaticSA} find that adding semantic facts can help LLM to improve performance on code summarization..~\citet{Wei2023CoeditorLC} propose a new model named Coeditor, which predicts edits to a code region based on recent changes within the same codebase using a multi-round code auto-editing setting.
In addition, introducing experience information such as historical interaction information can help the LLM-based agents better understand the context and make correct decisions. Some work uses experience information from past reasoning and decision-making processes to obtain more accurate answers by iteratively querying and modifying answers. For example,~\citet{Ren2023FromMT} propose KPC, a novel Knowledge-driven Prompt Chaining-based code generation approach, which decomposes code generation into an AI chain with iterative check-rewrite steps~\citet{Zhang2023RepoCoderRC} propose RepoCoder, a simple, generic, and effective framework which makes effective utilization of repository-level information for code completion in an iterative retrieval generation pipeline.  
\citet{Eghbali2024DeHallucinatorIG} present De-Hallucinator, an LLM-based code completion technique that retrieves suitable API references and iteratively queries the model with increasingly suitable context information in the prompt to ground the predictions of a model.

\subsubsection{Procedural Memory}
The procedural memory of Agents in software engineering contains the implicit knowledge stored in the LLM weights and the explicit knowledge written in the agent’s code.

\textbf{Implicit knowledge} is stored in the LLM parameters. Existing work usually proposes new LLMs with rich implicit knowledge to complete various downstream tasks, by training the model with a large amount of data.~\citet{zheng2023survey} sorted out the code LLMs in the SE field based on their affiliation type, including companies, universities, research teams\&open-source communities, and individuals\&anonymous contributors.

\textbf{Explicit knowledge} is written in the agent's code, enabling the agent to operate automatically. Several works~\citet{Patel2023EvaluatingIL, Shin2023PromptEO, Zhang2023PromptEnhancedSV} have explored different ways of constructing the agent’s code. Specifically,~\citet{Patel2023EvaluatingIL} use three types of in-context supervision to specify library functions including Demonstrations, Description, and Implementation.~\citet{Shin2023PromptEO} investigate the effectiveness of state-of-the-art LLM with three different prompting engineering techniques (i.e., basic prompting, in-context learning, and task-specific prompting) against fine-tuned LLMs on three typical ASE tasks.~\citet{Zhang2023PromptEnhancedSV} explore the performance of software vulnerability detection using ChatGPT with different prompt designs(i.e., Basic Prompting, Prompting with Auxiliary Information, and Chain-of-Thought Prompting).

\subsection{Action}
\label{sec:action}
The action module includes two types: internal and external actions. The external actions interact with the external environment to obtain feedback information, including Dialogue with humans/agents and interaction with the digital environment, while the internal actions reason and make decisions based on the input of the LLM and refine the decision based on the obtained feedback, including reasoning, retrieval, and learning actions. 
Next, we will introduce each action in detail.

\subsubsection{Internal Action}
Internal actions include reasoning, retrieval, and learning actions. Separately, reasoning actions are responsible for analyzing problems, reasoning, and making decisions based on the input of the LLM-based agent. Retrieval actions can retrieve relevant information from the knowledge base to assist reasoning actions in making correct decisions. Learning actions are continuously learning and updating knowledge by learning and updating semantic, procedural, and episodic memories, thereby improving the quality and efficiency of reasoning and decision-making.

\begin{figure*}[t]
    \centering
    \includegraphics[width=0.92\linewidth]{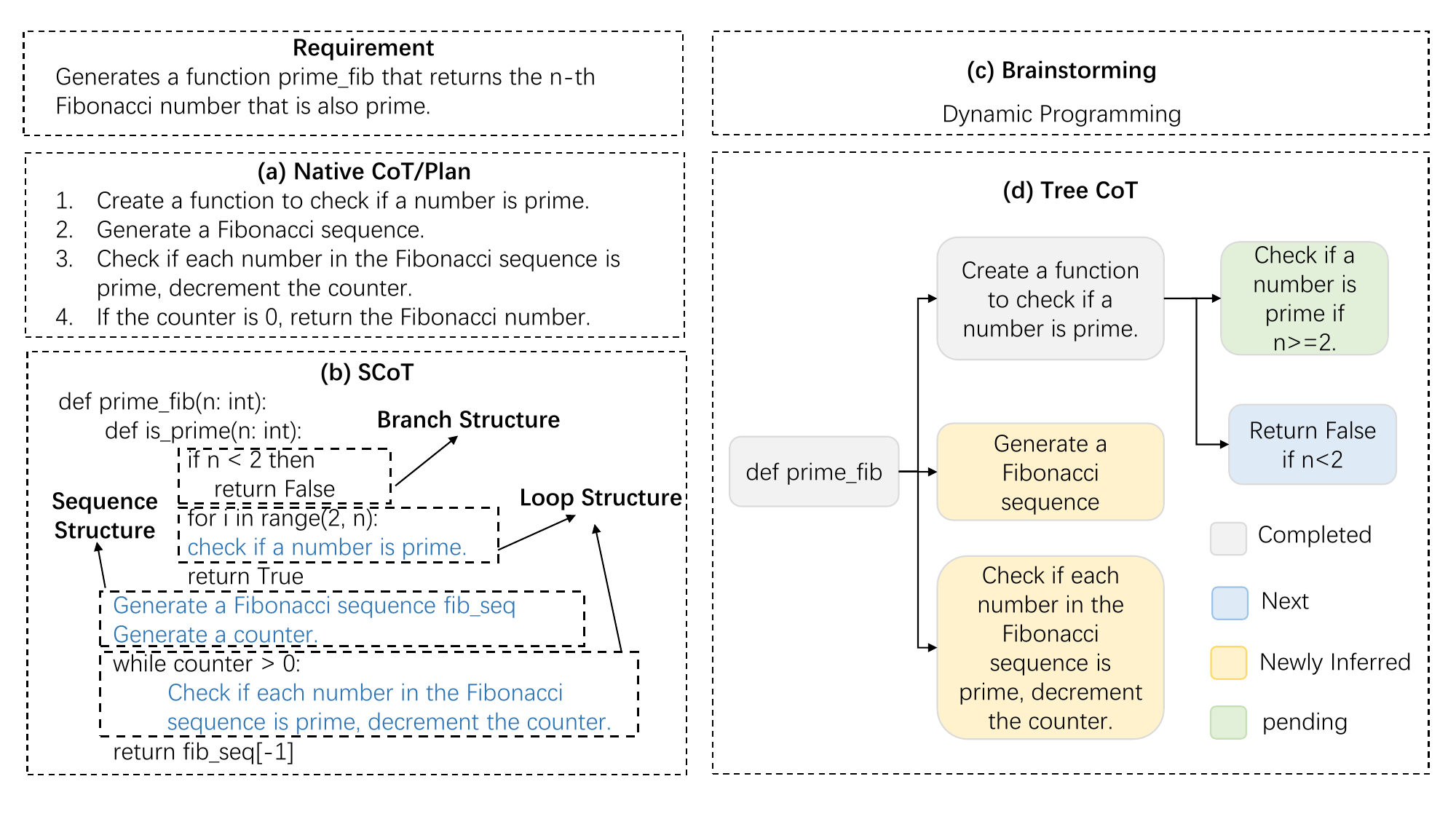}
    \caption{Different CoTs from different methods, where (a) is native cot/plan, which is obtained by letting LLM think step by step in the prompt and includes the detailed process of analyzing the problem and steps to solve the problem. (b) is Structured CoTs (SCoT), which combines code features to generate a code skeleton containing structures such as brach and loop in the graph. The blue font abstractly summarizes the description of LLM generating specific code based on SCoT. (c) is the result of brainstorming, which is obtained by analyzing the problem description and using knowledge of algorithms, data structures, and mathematics to provide ideas for solving it. (d) is an example of a Tree CoT, which dynamically explores and iteratively updates the CoT to gradually decompose and complete the problem.}
\label{fig:cot}
\end{figure*}

\textbf{Reasoning Action.}
A rigorous reasoning process is the key to completing tasks by LLM-based agents and Chain-of-Though (CoT) is an effective way of reasoning. With the help of CoT, the LLMs can deeply understand the problem, decompose complex tasks, and generate high-quality answers. As shown in Figure~\ref{fig:cot}, existing work has explored different forms of CoT, including naive CoT/Plan, SCoT,  brainstorming, tree CoT, etc. Specifically, naive CoT/Plan refers to a paragraph of text in the prompt describing the process of reasoning for the problem.
In the early work, a simple sentence was added to the prompt to guide LLMs in generating a chain of thought and better solving the problem. For example, ~\citet{Hu2024LeveragingPD} propose an in-context learning approach that uses a "print debugging" method to guide LLMs to debug. As LLM technology develops, the design of CoT has become more complex. Inspired by the process of developers validating the feasibility of test scenarios,~\citet{inproceedings} 
design chain-of-thought (CoT) reasoning to extract human-like knowledge and logical reasoning from LLMs. 
~\citet{Le2023CodeChainTM} propose CodeChain, a novel framework for inferences that generates a chain of self-revisions guided by some representative sub-modules generated in previous iterations.
~\citet{huang2024codecot} present Code Chain-of-Thought (CodeCoT) that generates test cases to validate whether the code has syntax errors during the execution and then employs a self-examination phase, integrating CoT with a self-examination process for code generation. \citet{Tian2023TestCaseDrivenPU} propose a novel prompting technique to devise both sophisticated thought-eliciting prompting and feedback based on prompting and make the first exploration to improve the code generation performance of LLMs.

Considering the characteristics of code, some works proposed structured CoT to introduce the structural information of the code. As shown in (b) in Figure ~\ref{fig:cot}, the structured CoT presents the reasoning process in a pseudo-code-like form, involving structures such as loops, branches, etc. For example,~\citet{Li2023StructuredCP} propose Structured CoTs (SCoTs) which can efficiently use the rich structural information of source code and present SCoT prompting, a novel prompting technique for code generation.
~\citet{Christianos2023PanguAgentAF} presents a general framework model to utilize the construction of intrinsic and extrinsic functions to add previous understandings of reasoning structures, integrating and learning structured reasoning into AI agents' policies.
In addition, some works have proposed other forms of CoT, such as brainstorming and tree-shaped CoT, as shown in (c) and (d) in Figure ~\ref{fig:cot}. Brainstorming is to generate related keywords based on the input. For example,~\citet{Li2023ThinkOT} introduces a novel Brainstorm framework for code generation which leverages a brainstorming step that generates and selects diverse thoughts on the problem to facilitate algorithmic reasoning before generating code. The tree-shaped CoT~\cite{Feng2023PromptingIA} dynamically explores and updates the CoT, and the nodes in the tree involve multiple states including completed, new, and newly inferred, pending. 

There are also some studies exploring other techniques to improve the reasoning ability and reasoning efficiency of LLM-based agents. For example,
~\citet{Wang2024TeachingCL} propose TOOLGEN that comprises Trigger Insertion and Model Fine-tuning phases (Offline), and Tool-integrated Code Generation phases (Online). TOOLGEN  reasons the positions to trigger auto-completion tools using the augments functions within a given code corpus with a special mark token.
~\citet{Yang2023ChainofThoughtIN} design a novel approach COTTON which can leverage lightweight Language Models to generate CoTs for code generation automatically.
~\citet{Zhang2023DraftV} present self-speculative decoding, a novel inference scheme that generates draft tokens and then employs the original LLM to validate those draft output tokens in one forward pass.
~\cite{zhou2023adaptive} introduce an Adaptive-Solver framework that strategically adjusts the solving strategy according to the difficulty of the problem, which not only improves the computational efficiency but also improves the overall performance.

\textbf{Retrieval Action.}
The retrieval action can retrieve relevant information from the knowledge base to assist the reasoning action in making correct decisions. The input used for retrieval and the output content obtained by retrieval have different types. As shown in Table~\ref{tab:retrieval}, the input and output can be text, code, or hybrid information containing both text and code. Specifically, it can be divided into the following types: 
\textbf{(1) Text-Code.} Typically requirements are treated as input to retrieve related code or used APIs which are added to the prompt to generate responding code. For example,~\citet{Zan2022WhenLM} propose a novel framework with APIRetriever and  APICoder modules. 
Specifically, the APIRetriever retrieves useful APIs, and then the APICoder generates code using these retrieved APIs. De-Hallucinator~\cite{Eghbali2024DeHallucinatorIG} retrieves suitable API references adding to the prompt and iteratively querying the model with the obtained prompt.
\textbf{(2) Text-Text.} Sometimes, requirements are also used as input to retrieve relevant documentation or similar questions to help complete the task. For example,~\citet{Zhou2022DocPromptingGC} introduce DocPrompting, a natural language-to-code generation method that explicitly leverages documents by retrieving relevant document fragments for a given NL intent. ~\citet{Zhang2024CodeAgentEC} present CodeAgent, a novel LLM-based agent framework that integrates external tools to retrieve relevant information for effective repo-level code generation, enabling interaction with software artifacts for information retrieval, code symbol navigation, and code testing.
\textbf{(3) Code-Code.} Code can also be used as input to retrieve similar or related code to provide a reference for generating the target code. For example,~\citet{Zhang2023RepoCoderRC} propose RepoCoder, a simple, generic, and effective framework that retrieves similarity-based repository-level information in an iterative retrieval generation pipeline.
\textbf{(4) Hybrid-Code.} In addition to using a single type of information such as text or code as input to retrieve related code, multiple types of information can also be combined into hybrid information to improve retrieval accuracy. For example,~\citet{Li2022GenerationAugmentedQE} utilizes the powerful code generation model to benefit the code retrieval task by augmenting the documentation query with its generated code snippets from the code generation model (generation counterpart) and then uses the augmented query to retrieve code. ToolCoder~\cite{Zhang2023ToolCoderTC} uses an online search engine and documentation search tool to get the proper API suggestion, assisting in API selection and code generation.
In addition, the retrieved content is not limited to a single type of information.
\textbf{(5) Code-Hybrid.} It uses code as input and retrieves various relevant information. For example,~\citet{Nashid2023RetrievalBasedPS} presents a novel technique named CEDAR for prompt creation that automatically retrieves code demonstrations similar to the code-related developer task, based on embedding or frequency analysis.~\citet{Geng2023LargeLM} generates multi-intent comments for code by adopting the in-context learning paradigm which selects various code-comments examples from the example pool.
\textbf{(6) Text-Hybrid.} It uses requirements as input to retrieve relevant code and similar questions to be referenced. For example,~\citet{Li2023LargeLM} propose LAIL (LLM-Aware In-context Learning), a novel learning-based selection approach to select examples for code generation.~\citet{Li2023AceCoderUE} introduces a novel mechanism named AceCoder which uses requirements to retrieve similar programs as examples in prompts, which provide lots of relevant content (e.g., algorithms, APIs). 

\begin{figure*}[t]
    \centering
    \includegraphics[width=0.92\linewidth]{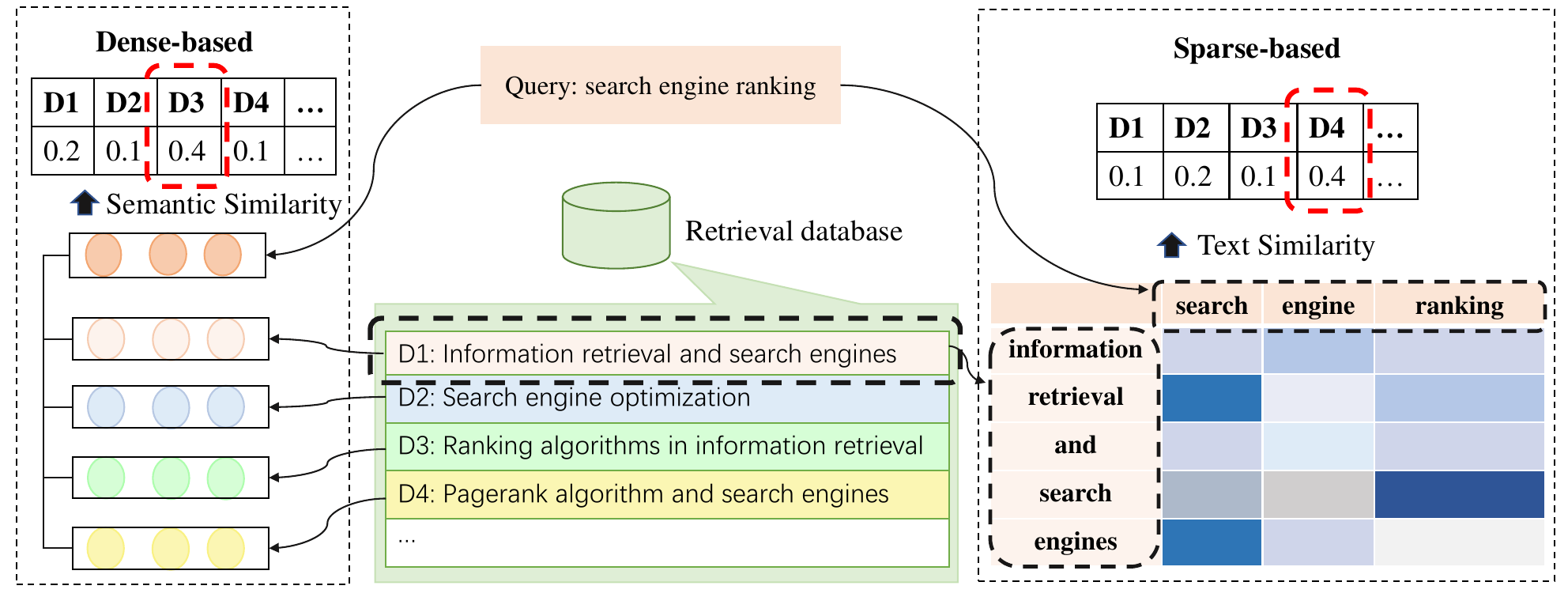}
    \caption{The pipeline of different retrieval methods. The left part is the pipeline of the dense-based retrieval method, which can use different models to convert the text into a high-dimensional embedding vector and compare the semantic similarity to retrieve the sample with the highest similarity score. The right part is the pipeline of the sparse-based retrieval method, which just compares the text similarity and ignores semantics.}
\label{fig:retrieval_method}
\end{figure*}

According to previous research~\cite{li2022survey, zhao2024retrieval, hu2023revisiting}, existing retrieval methods can be divided into sparse-based retrieval, dense-based retrieval~\cite{wang2024rlcoder}, and other methods~\cite{hayati2018retrieval, zhang2020retrieval, poesia2022synchromesh, ye2021rng, shu2022tiara}. Figure~\ref{fig:retrieval_method} shows the pipelines of sparse-based retrieval and dense-based retrieval respectively. The dense-based retrieval method converts the input into a high-dimensional vector and then compares the semantic similarity to select the k samples with the highest similarity, while the sparse-based retrieval method calculates metrics such as BM25 or TF-IDF to evaluate the text similarity between samples. Additionally, various alternative retrieval methods have been explored. For instance, some studies focus on calculating the edit distance between natural language texts~\cite{hayati2018retrieval} or abstract syntax trees (ASTs) of code snippets~\cite{zhang2020retrieval, poesia2022synchromesh}. And some approaches leverage knowledge graphs for retrieval~\cite{ye2021rng, shu2022tiara}.

Dense-based and sparse-based retrieval methods are the two most mainstream retrieval methods. Among them, dense-based retrieval techniques generally offer superior performance compared to sparse-based retrieval methods. However, sparse-based retrieval is often more efficient and can achieve comparable performance levels to dense-based retrieval. As a result, many studies choose to employ sparse-based retrieval methods due to their efficiency.

\begin{table*}[ht]
\caption{Different types of retrieval actions classified by input and output.}
    \centering
    \small
    \begin{tabular}{m{2cm}m{3cm}m{3.5cm}m{5.5cm}}
    \toprule
    Types &  Input & Output & Studies \\
    \midrule
    Text-Code &  Requirements & API & APIRetriever~\cite{Zan2022WhenLM}, \newline{De-Hallucinator~\cite{Eghbali2024DeHallucinatorIG}}\\
    \midrule
	Text-Text &Requirements &Relevant Documents, \newline{Repo Documentation}&DocPrompting~\cite{Zhou2022DocPromptingGC}, \newline{CodeAgent~\cite{Zhang2024CodeAgentEC}}\\
    \midrule
    \multirow{2}*{Code-Code} &Incomplete Code&Similar Code&RepoCoder~\cite{Zhang2023RepoCoderRC} \\
    &Target Hole's Location&Repo Context&RepoFusion~\cite{shrivastava2023repofusion}\\
    \midrule
    \multirow{2}*{Hybrid-Code} &Requirements, \newline{Unfinished Code}&API&ToolCoder~\cite{Zhang2023ToolCoderTC}\\
     &Requirements, \newline{Generated Code} &Target Code& Query Expansion~\cite{Li2022GenerationAugmentedQE}\\
    \midrule
    Code-Hybrid &Code Snippet&Code-Comment Examples&CEDAR~\cite{Nashid2023RetrievalBasedPS}, \newline{Multi-intent~\cite{Geng2023LargeLM}}\\
    
    \midrule
    
    Text-Hybrid  &Requirements&Examples&LAIL~\cite{Li2023LargeLM}, \newline{AceCoder~\cite{Li2023AceCoderUE}}\\
    \bottomrule
\end{tabular}
\label{tab:retrieval}
\end{table*}

\textbf{Learning Action.}
Learning actions are continuously learning and updating knowledge by learning and updating semantic and procedural memories, thereby improving the quality and efficiency of reasoning and decision-making.
\textit{(1) Updating Semantic Memory with Knowledge.} Semantic memory mainly exists in the knowledge base that stores basic world knowledge and can be updated by updating the knowledge base using recognized code knowledge or constructing a new one. For example,~\citet{liao2023context} propose a novel code generation framework, called $A^{3}$-CodGen, which generates higher quality code by leveraging the information retrieved from a retrieval base that contains three types of information: local awareness information from the current code file, global awareness information from other code files, and third-party library information.~\citet{du2024vul} propose a novel LLM-based vulnerability detection technique Vul-RAG which constructs a vulnerability knowledge base by extracting multi-dimension knowledge via LLMs from existing CVE instances. 
\textit{(2) Updating Implicit Knowledge.} Since the implicit knowledge is stored in the LLM parameters, it can be updated by fine-tuning the model to update the LLM parameters. Early work usually constructs new data to supervise fine-tuning the pre-trained model which updates the full parameters of the model~\cite{Xia2023RevisitingTP, Wei2023CoeditorLC, Wei2023MagicoderSC, tao2024kadel, wang2024sparsecoder, liu2023refbert, wang2023you, shi2023cocosoda}. However, as the parameter scale increases, the cost of fine-tuning the model also increases. Some works try to explore parameter-efficient fine-tuning techniques~\cite{Weyssow2023ExploringPF, shi2023towards}. For example,~\citet{Weyssow2023ExploringPF} deliver a comprehensive study of Parameter-Efficient Fine-Tuning (PEFT) techniques(PEFT) techniques for LLMs under the automated code generation scenario.~\citet{Wang2023OneAF} insert and fine-tune the parameter-efficient structure adapter, rather than fine-tuning the pre-trained model. Most of the current work uses effective fine-tuning technologies to fine-tune the model~\cite{guo2024stop, shi2023sotana}.

\textit{(3) Updating Agent Code.} 
Agent code refers to the program or algorithm that the agent runs to guide its behavior and decision-making. The LLM-based agent constructs corresponding prompts as agent codes to regulate how to perceive the environment, reason and make decisions, and perform actions. Many works use instruction-tuning techniques to align the output of LLM with the input instructions. For example,.~\citet{Muennighoff2023OctoPackIT} leverage the natural structure of Git commits that pair code changes with human instructions and apply instruction tuning using them.~\citet{Hu2023InstructCoderEL} constructed the first instruction-tuning dataset named InstructCoder which is designed to adapt LLMs for general-purpose code editing. These high-quality quality data can bring new knowledge to the larger language model and update semantic memory.~\citet{Zan2023CanPL} conduct extensive experiments of 8 popular programming languages on StarCoder to explore whether programming languages can boost each other via instruction-tuning.

\subsubsection{External Action}

\textbf{Dialogue with Human/Agents}
Agents can interact with humans or other agents, and get rich information in the interaction process as feedback, expanding the knowledge of the agent and refining the answers of LLM more corrector. Specifically, many works use LLMs as agents to interact~\cite{lu2024yoda, Jain2023CoarseTuningMO, Paul2023REFINERRF, Shojaee2023ExecutionbasedCG, Liu2023RLTFRL, Wang2023ChatCoderCR, Mu2023ClarifyGPTEL, Madaan2023SelfRefineIR}.
~\citet{Jain2023CoarseTuningMO} propose RLCF that uses feedback from a different LLM that compares the generated code to a reference code to further train a pre-trained LLM via reinforcement learning. REFINER~\cite{Paul2023REFINERRF} is a framework that can interact with a critic model that provides automated feedback on the reasoning.~\citet{Yang2023SupervisedKM} study and elucidate how LLMs benefit from discriminative models.~\citet{Moon2023CoffeeBY} construct a new dataset specifically designed for code fixing with feedback and then use this dataset to gain a model that can automatically generate helpful feedback for code editing via Preference-Optimized Tuning and Selection. PPOCoder~\cite{Shojaee2023ExecutionbasedCG} consists of two parts, critic and actor, and will be optimized with PPO through the interaction between these two models. RLTF~\cite{Liu2023RLTFRL} interacts with other models that utilize ground truth data and online buffer data generated by interacting with the compiler to calculate loss and update the model weights through gradient feedback.~\citet{Wang2023ChatCoderCR} propose ChatCoder, a method to refine the requirements via chatting with large language models.~\citet{Sun2023CloverCV} propose Clover that lies a checker to check the consistency among code, docstrings, and formal annotations. ClarifyGPT~\cite{Mu2023ClarifyGPTEL} prompts another LLM to generate targeted clarifying questions to refine the ambiguous requirement inputted by users. Reflexion~\cite{Shinn2023ReflexionLA} can interact with humans and other agents to generate external feedback. Self-Refine~\cite{Madaan2023SelfRefineIR} uses a single LLM as the generator, refiner, and feedback provider, rather than requiring any supervised training data, additional training, or reinforcement learning. Repilot~\cite{Wei2023CopilotingTC} synthesizes a candidate patch through the interaction between an LLM and a Completion Engine. Specifically, Repilot prunes away infeasible tokens suggested by the LLM.~\citet{Wang2023MINTEL} introduce MINT, a benchmark that can evaluate LLMs' ability to solve tasks with multi-turn interactions by leveraging users' natural language feedback simulated by GPT-4.\citet{Hong2023MetaGPTMP} propose MetaGPT, an innovative meta-programming framework that incorporates efficient human workflows into LLM-based multi-agent collaborations.~\citet{Huang2023AgentCoderMC} introduces AgentCoder, a novel code generation solution comprising a multi-agent framework with specialized agents: the programmer agent, the test designer agent, and the test executor agent.

\textbf{Digital Environment}
Agents can interact with digital systems, such as OJ platforms, web pages, compilers, and other external tools, and the information obtained during the interaction process can be used as feedback to optimize themselves.
Specifically, compilers are the most common external tools~\cite{Jain2023CoarseTuningMO, Shojaee2023ExecutionbasedCG, Liu2023RLTFRL, Wang2022CompilableNC, Zhang2023SelfEditFC}. For example, 
RLCF~\cite{Jain2023CoarseTuningMO} trains the pre-trained LLM via reinforcement learning, using the compiler-derived feedback on whether the code it generates passes a set of correctness checks. PPOCoder~\cite{Shojaee2023ExecutionbasedCG} can incorporate compiler feedback and structure alignments as extra knowledge into the model optimization to fine-tune code generation models via deep reinforcement learning (RL). 
RLTF~\cite{Liu2023RLTFRL} interacts with the compiler to produce a training data pair and then stores it in the online buffer. 
~\citet{Wang2022CompilableNC} propose COMPCODER that utilizes compiler feedback for compilable code generation.
~\citet{Zhang2023SelfEditFC} propose Self-Edit, a generate-and-edit approach that utilizes execution results of the generated code from LLMs to improve the code quality on the competitive programming task.
In addition, many works construct tools such as search engines, completion engines, and others to expand the capabilities of intelligent agents~\cite{Wang2024TeachingCL, Zhang2024CodeAgentEC, Agrawal2023GuidingLM, Wei2023CopilotingTC, zhang2023code}.
~\citet{Wang2024TeachingCL} introduce TOOLGEN, an approach that integrates autocompletion tools into the code LLM generation process to address dependency errors such as undefined-variable and no-member errors.~\citet{Zhang2024CodeAgentEC} present CodeAgent, a novel LLM-based agent framework that integrates five programming tools, enabling interaction with software artifacts for information retrieval, code symbol navigation, and code testing for effective repo-level code generation.~\citet{Agrawal2023GuidingLM} propose MGD, a monitor-guided decoding method where a monitor uses static analysis to guide the decoding. Repilot~\cite{Wei2023CopilotingTC} synthesizes a candidate patch through the interaction between an LLM and a Completion Engine. Specifically, Repilot completes the token based on the suggestions provided by the Completion Engine.

\section{Challenges and Opportunities}
\label{sec:challenges&opportunities}
Upon analyzing the work related to LLM-based agents in software engineering, it is evident that there are still many challenges in the current integration of these two fields, which limits the development of both. In this section, we will discuss in detail the challenges faced by current LLM-based agents in SE and discuss some opportunities for future work based on the analysis of existing challenges.

\subsection{Lack of Exploration of Perception Module}
As mentioned in Section~\ref{sec:perception}, there is a lack of work exploring the perception module of LLM-based agents in SE. Unlike natural language, code is a special representation that can be treated as ordinary text or converted into an intermediate representation with code characteristics, such as AST, CFG, and so on. Existing works~\cite{ahmed2024automatic, alkaswan2023extending, arakelyan2023exploring, Beurer_Kellner_2023, Alqarni2022LowLS} often treat code as text, and there is still a lack of work on LLM-based agents in SE that explore tree/graph-based input modalities. In addition, there is still a lack of research on exploring visual and auditory input modalities. 

\subsection{Role-playing Abilities}
LLM-based agents are often required to play different roles across a range of tasks, necessitating specific skills for each role. For example, an agent may function as a code generator when tasked with generating code, and as a code tester when tasked with code testing. Furthermore, in certain scenarios, these agents may need to have multiple capabilities at the same time. For example, in the code generation scenario, the agent needs to play the role of both a code generator and tester and accordingly needs to have the ability to both generate and test code~\cite{huang2023agentcoder}. In the software engineering field, there are various niche tasks for which LLM learning is not enough and complex tasks that require agents to have multiple capabilities, such as test generation scenario, front-end development, repository-level issue resolution, etc. \textit{Therefore, advancing research on how to enable agents to effectively adopt new roles and manage the demands of multi-role performance represents a promising direction for future work.}

\subsection{Lack of Knowledge Retrieval Base}
The external knowledge retrieval base is an important part of the semantic memory in the agent memory module and one of the important external tools that the agent can interact with. In NLP fields, there are knowledge bases such as Wikipedia as external retrieval bases~\cite{zhao2023verify}. However, in the SE field, there is currently no authoritative and recognized knowledge base that contains rich code-related knowledge, such as the basic syntax of various programming languages, various commonly used algorithms, knowledge related to data structures and operating systems, etc. \textit{In future research, efforts could be directed towards developing a comprehensive code knowledge base to serve as an external retrieval base for the agent. This knowledge base would enrich the available information, thereby enhancing both the quality and efficiency of reasoning and decision-making processes.}

\subsection{Hallucinations of LLM-based Agents}
Many studies related to LLM-based agents consider LLMs as the cognitive core of the agents, with the agents' overall performance being closely tied to the capabilities of the underlying LLMs. Existing research~\cite{pan2024lost,liu2024exploring} has shown that LLM-based agents may produce hallucinations, such as generating non-existent APIs when completing tasks in the SE field. Mitigating these hallucinations can improve the overall performance of the agents. At the same time, the optimization of the agents can also reversely alleviate the hallucinations of the LLM-based agents, highlighting the bidirectional relationship between agent performance and hallucination mitigation. Although some efforts have been made to investigate the hallucinations of LLMs, significant challenges remain in addressing the hallucination issue within LLM-based agents. \textit{Exploring what types of hallucinations exist in LLM-based agents, deeply analyzing the causes of these hallucinations, and proposing effective hallucination mitigation methods are great opportunities for future research.}

\subsection{Efficiency of Multi-agent Collaboration}
In the process of multi-agent collaboration, each individual agent needs to play different roles to accomplish specific tasks, and the outcomes of each agent's decisions are then combined to collectively tackle more complex objectives~\cite{chen2023agentverse, hong2023metagpt, huang2023agentcoder, wang2023mac}. However, this process often requires a large amount of computing resources for each agent, resulting in resource waste and reduced efficiency. In addition, each single agent needs to synchronize and share various types of information, which introduces additional communication costs and affects the real-time and response speed of collaboration. \textit{Effectively managing and allocating computing resources, minimizing inter-agent communication costs, and reducing the reasoning overhead of individual agents represent key challenges for enhancing the efficiency of multi-agent collaboration. Addressing these issues presents a significant opportunity for future research.}

\subsection{SE Technologies in LLM-based Agents}
\label{sec:se4agent}
Techniques from the software engineering domain, particularly those to code, have the potential to significantly advance the development of the agent field, indicating a mutually beneficial relationship between these two domains.
For example, software testing techniques can be adapted to identify abnormal behaviors and potential defects in LLM-based agents. Additionally, improvements in software tools, such as APIs and libraries, can also boost the performance of LLM-based agents especially for those with tool using abilities. 
Furthermore, software package management techniques can be adapted for effectively managing agent systems. For example, version control can be applied to monitor and coordinate updates across different agents in an agent system, enhancing compatibility and system integrity.  
 
However, research in this line remains limited. 
\textit{Therefore, exploring the incorporation of more sophisticated SE techniques into agent systems represents a promising area for future research, with the potential to drive advancements in both fields.}

\section{Conclusion}
To conduct an in-depth analysis of the work on combining LLM-based agents with SE, we first collect many studies that combine LLM-based agents with tasks in the software engineering field. Then, we present a framework of LLM-based agents in software engineering which contains three key modules: perception, memory, and actions, after sorting out and analyzing the studies obtained during data collection. Finally, we introduce the details of each module in the framework, analyze the current challenges for LLM-based agents in the SE field, and point out some opportunities for future work.

\bibliography{anthology,custom}
\bibliographystyle{acl_natbib}

\end{document}